\documentclass{mn2e}
\def \th {\thinspace}
\def \src {XB\thinspace 1746-371}
\def \degmark{^\circ}

\def\approxgt{\mathrel{\hbox{\rlap{\lower.55ex \hbox {$\sim$}}
\kern-.3em \raise.4ex \hbox{$>$}}}}
\def\approxlt{\mathrel{\hbox{\rlap{\lower.55ex \hbox {$\sim$}}
\kern-.3em \raise.4ex \hbox{$<$}}}}
\def \th {\thinspace }
\def \ref {\reference{}}

\def \degmark{^\circ}

\usepackage{graphicx}

\title[The orbital period of XB\th 1746-371]             
{The orbital period of the dipping, bursting, globular cluster X-ray source XB\th 1746-371
from Rossi X-ray Timing Explorer observations}

\author[Ba\l uci\'nska-Church, Church \& Smale]    
{M. Ba\l uci\'nska-Church$^{1,2}$, M. J. Church$^{1,2}$
\newauthor
and A. P. Smale$^{3,4}$\\
      $^1$University of Birmingham, School of Physics and Astronomy,
      Birmingham, B15 2TT, UK\\
      $^2$Astronomical Observatory, Jagiellonian University, ul. Orla 171,
      30-244 Cracow, Poland\\
      $^3$Laboratory for High Energy Astrophysics, Code 660.1, NASA/Goddard Space 
      Flight Center, Greenbelt, MD 20771\\
      $^4$Office of Space Science, Astronomy and Physics Division, Code SZ,
      NASA Headquarters, Washington, DC 20546-0001}

\date{Accepted 2003 September 18. Received 2003 July 8}

\begin{document}
\maketitle


\begin{abstract}

We present results from two long observations of XB\th 1746-371 by the
{\it Rossi X-ray Timing Explorer (RXTE)} in 2002 January and May,
lasting 4 and 5 days respectively. Dips are observed in the X-ray
light curves with a depth of 25 per cent, largely independent of energy
within the usable band of the PCA instrument of 2.1 -- 16.0 keV. X-ray
bursting and flaring activity are also evident. The dips define the
orbital period of the system, and using a power spectral analysis and
a cycle counting technique we derive an accurate period of
$P_{orb}$=5.16$\pm$0.01 hr. The previously-reported candidate period
of 5.73$\pm$0.15 hr, obtained using {\it Ginga} data, is inconsistent
with our determination, perhaps due to the weakness of the dipping and
the variability of the source during that observation. The dips in the
{\it RXTE} observations presented here do not align with the {\it
Ginga} period, however our improved period is consistent with a wide
range of archival data.

\end{abstract}


\section{Introduction}

XB\th 1746-371 is one of the group of $\sim$10 low mass X-ray binaries (LMXB)
exhibiting X-ray dipping at the orbital period, generally accepted as being due to absorption in
the bulge in the outer accretion disc. These sources provide substantially more diagnostics
and information on the geometry, size and properties of the X-ray emission regions than
non-dipping sources. Firstly, there is strong, complex, spectral
evolution in dipping; 
a successful emission model must be able to realistically fit spectra
selected in intensity bands through the dips, as
well as the non-dip spectrum. In recent years, all
of the dipping sources have been fitted by a model consisting of pointlike blackbody
emission from the surface of the neutron star which causes fast variability in dipping, plus
Comptonized emission from an extended accretion disc corona (ADC) 
(Ba\l uci\'nska-Church et al. 1999, 2000; Church et al. 1997, 1998a, 1998b; 
Smale, Church \& Ba\l uci\'nska-Church 2001, 2002). Secondly,
the technique of dip ingress timing allows measurement of the size of extended emission
regions, when overlapped by an absorber of larger angular size. This reveals
that the ADC is very large, of radial extent $r_{\rm ADC}$ typically 50,000 km, and that 
$r_{\rm ADC}$ increases with source luminosity (Church \& Ba\l uci\'nska-Church 2003)
such that in the brightest dipping source X\th 1624-490, $r_{\rm ADC}$ becomes
700,000 km, or 65 per cent of the radius of the accretion disc. Thirdly, long-term monitoring
of the light curves of the dipping sources may reveal evolution in the orbital period, or 
other effects. This is the case in the source XB\th 1916-053 having an X-ray period
of $\sim$3000 s, 1 per cent shorter than the optical period (Chou, Grindlay \& Bloser 2001;
Homer et al. 2001). 
In the case of XBT\th 0748-676, timing studies have been aided by the
observation of almost complete X-ray eclipses (Parmar et al. 1986) by the companion 
star. 

XB\th 1746-371 is a medium brightness member of the dipping LMXB class in the core of the globular
cluster NGC\th 6441 (Giacconi et al. 1974; Clark et al. 1974; Grindlay et al. 1976; 
Jernigan \& Clark 1979; Hertz \& Grindlay 1983). 
Dipping was discovered by Parmar, Stella \& Giommi (1989) in an {\it Exosat} observation
indicating a period 5.0$\pm$0.5 hr. These dips had a depth of only
15 per cent in the band 1 -- 10 keV, and X-ray bursts were also detected (Sztajno et al. 1987).
Sansom et al. (1993) used {\it Ginga} data to find an orbital period of 5.73$\pm$0.15 hr.
More recently, Jonker et al. (2000) discovered a 1 Hz quasi-periodic
oscillation, present during persistent emission and bursts,
using {\it RXTE} data. An observation with {\it BeppoSAX}
confirmed that dipping was apparently energy-independent as seen in {\it Exosat} 
(Parmar et al. 1999), as is also the case in another dipping source X\th 1755-338
(White et al. 1984; Mason, Parmar \& White 1985; 
\hbox{Church \& Ba\l uci\'nska-Church 1993).} 
\begin{figure*}                                                  
\begin{center}
\includegraphics[width=40mm,height=174mm,angle=270]{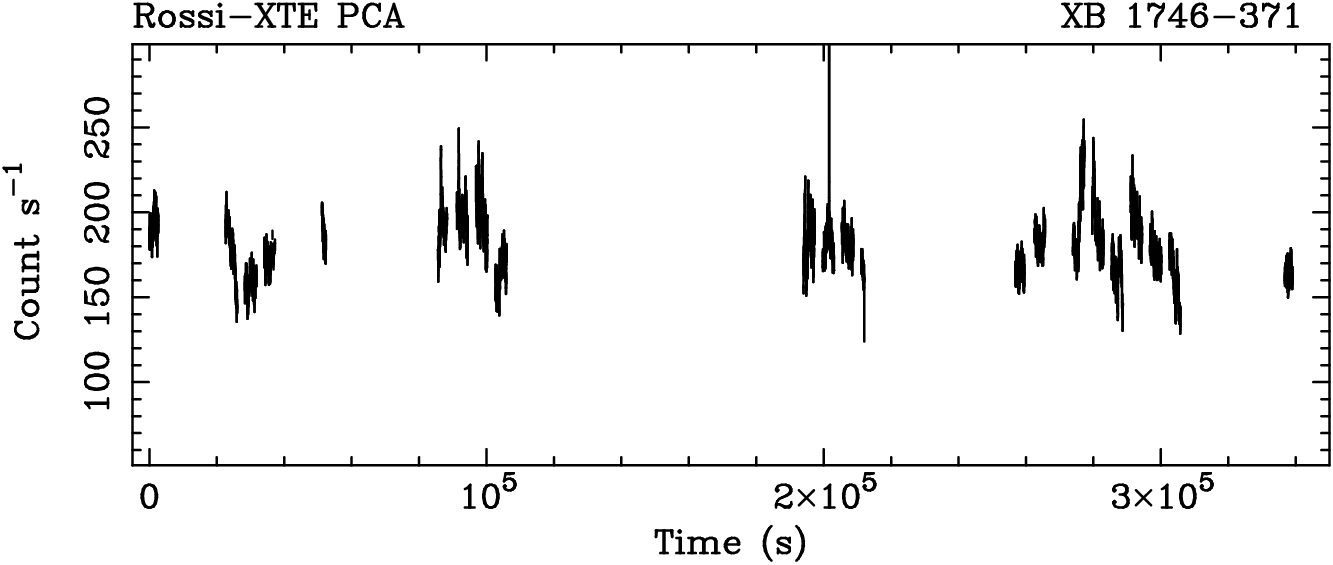}   
\includegraphics[width=40mm,height=174mm,angle=270]{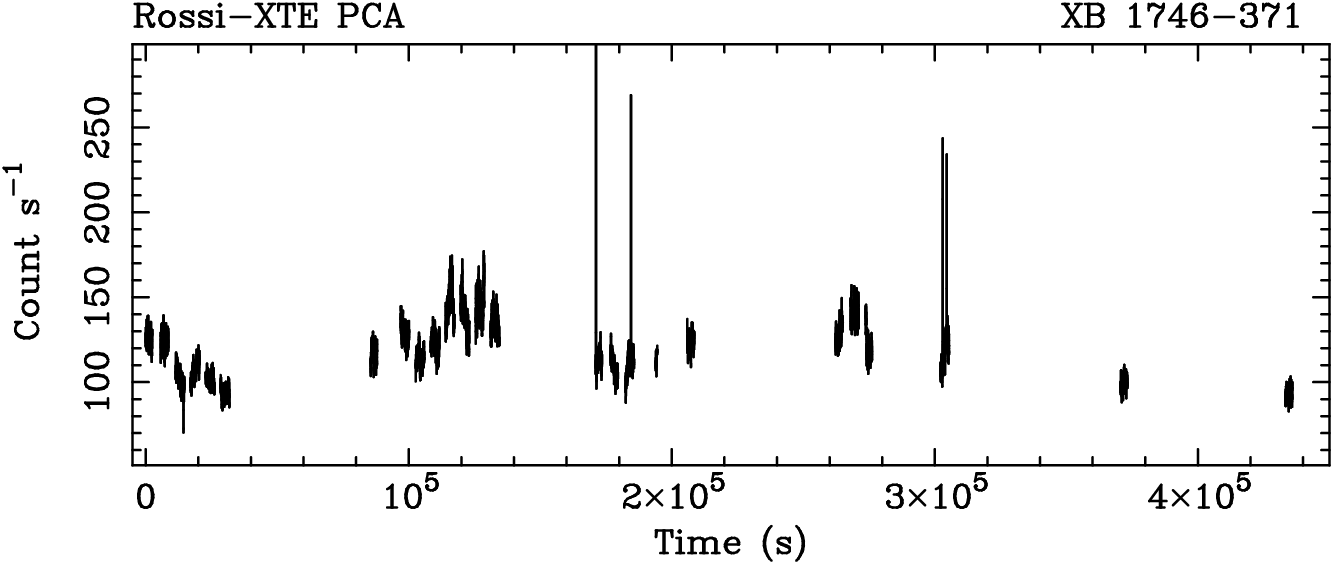}   
\caption{Background-subtracted, PCA light curves of the two observations in PCU0 and
PCU2 with 64 s binning in the total usable energy band 2.1 -- 16.0 keV.
Upper panel: 2002, January 16 -- 20; lower panel: 2002, May 3 -- 8.}
\end{center}
\end{figure*}
The {\it BeppoSAX} data implied an orbital period of $5.8^{+0.3}_{-0.9}$ hr.

In this paper we derive an orbital period for 
XB\th 1746-371 using {\it RXTE} observations performed in 2002 January 
and May, and refine this period 
using prior {\it RXTE} observations made in 1996 October
and 1998 June -- November.
We also re-examine archival data from {\it Exosat} 
and {\it RXTE} and show that the new period is consistent with these observations.
Our period is inconsistent with the {\it Ginga} period, and is closer to the original determination
by Parmar et al. (1989). In a further paper, we will present results for spectral
evolution during dipping and flaring.

\section{Observations and analysis}

We observed \src\ with  {\it RXTE} (Bradt et al. 1993) from 2002, January 16 UT 22:42:35
to January 20 UT 21:45:15 (observation ID 60044-02-01--00; and from
May 3 UT 14:22:08 to May 8 UT 16:15:15 (60044-02-02-00). The observations span nearly 
4 and 5 days, respectively. Results presented here
use data from the Proportional Counter Array (PCA) operating in Standard 2
mode with 16 s time resolution.  The PCA consists of five Xe proportional counter units,
numbered 0--4, with a combined effective area of about 6500 cm$^2$ (Jahoda et al. 1996).
PCUs 0 and 2 were reliably on during most of the observations, and we use these PCUs
in the following. The propane layer 
of PCU0 had failed at the start of epoch 5 (defined as 2000, May 13), however, this does not 
appreciably affect the light curves presented here. Light curves and spectra were 
extracted using the standard {\it RXTE} analysis software {\sc ftools 5.2}. 
Standard screening
was applied to select only data with an offset of the telescope pointing axis
from the source less than 0.02$\degmark$, and an elevation above the Earth's limb $>$ 10$\degmark$.
The source was bright during both observations, with count rates exceeding 40 c s$^{-1}$ 
per PCU, and thus the background was calculated using the program {\sc pcabackest}
with the latest version of the ``bright'' background model for epoch 5
appropriate to the observations released in January, 2002.
The energy band used, 2.1 -- 16.0 keV, was selected by comparison of the spectra of the source + background
and the background. At 16 keV the background was two times smaller than the source + background,
and data were not used at energies greater than this, where the spectra
were converging due to the increase of background with energy.

\section{Results}
    
The background subtracted light curves of the 2002 January and May observations are shown 
in Fig. 1.
Both observations reveal dipping with an intensity reduction
of 25 per cent. A total of 4 X-ray bursts can be seen, one of which is double.
The peak height of the bursts is reduced substantially by the binning which is comparable 
with the total burst duration.
In addition, there are periods of X-ray flaring in which the intensity
increases by 30 per cent for several thousand seconds. 
Inspection of
light curves in various sub-bands confirmed that dipping was evident in all energy bands,
whereas flaring was only visible above $\sim$5 keV.
Flaring is well-known in the bright
Z-track LMXB sources that are approaching the Eddington limit. However, the luminous
dipping source X\th 1624-490 also exhibits strong flaring 
dominating the light curve above 8 keV, and the spectral evolution
suggests marked changes on the neutron star and at the inner disc (Ba\l uci\'nska-Church et al. 2001).
We have also detected similar flaring in the dipping source XB\th 1254-490 (Smale et al. 2002).
In general, X-ray bursting is observed in faint sources, and X-ray flaring
in bright sources, so observation of both in the present observation is unusual
and presumably reflects the intermediate luminosity of the source. Spectral fitting a
non-dip, non-burst, non-flare spectrum gave a 1 -- 30 keV luminosity of
$1.36\times 10^{37}$ erg s$^{-1}$ for a distance of 9.0 kpc (Christian \& Swank 1997).
This is indeed intermediate when compared with X\th 1624-490 (above) with
a non-flaring luminosity of ${\rm 7\times 10^{37}}$ -- ${\rm 1.2\times 10^{38}}$ erg s$^{-1}$
in {\it BeppoSAX} and {\it RXTE} observations (Ba\l uci\'nska-Church et al. 2000, 2001)
(for a distance of 15 kpc Christian \& Swank 1997), 
and with typical burst sources such as XB\th 1323-619
(Barnard et al. 2001) with luminosities of a few $10^{36}$ erg s$^{-1}$.

The orbital period in XB\th 1746-371 was obtained by a two-stage
process. In the first stage, 
we obtained the power spectrum of the light curve for each
observation in the band 2.1 -- 3.7 keV.
In this band dipping was as strong as in any other band, but flaring was absent, so
that spurious peaks in the power spectrum were avoided.
The X-ray bursts were also removed from the data to prevent these contaminating the power spectrum.
Next, a technique was used to prevent spurious peaks in the power spectrum 
due to the data gaps caused by Earth occultation and South Atlantic Anomaly (SAA) passage.
A modified light curve was made in which real data was given an intensity of unity, and
zeroes written into the data gaps. From this, a window function of the data gaps was obtained
which allowed the cleaned spectrum to be obtained by deconvolution of
the uncleaned power spectrum and the window function (Roberts, Leh\'ar \& Dreher 1987).
This was done using the {\sc clean} alogorithm of the 
{\sc period} program which is part of the
{\sc starlink} packages. The incomplete coverage of individual dips did not justify
modelling the dip shape to establish the dip centre more accurately.
The data were also detrended, although this had little effect on the results.

In principle, a period determination might be made by fitting both
observations simultaneously. However, it can be seen from Fig. 1 that the January observation
constrains the orbital period well, as there are two clear dips towards the start of the
observation, and a clear dip close to the end, whereas the second observation only has clear
dipping at the start, and these data do not constrain the orbital period well. We therefore
obtain the orbital period from the first observation, and 
then confirm that this period
predicts dipping at times observed in the second observation (below).
The power spectrum of the first observation is presented in Fig. 2. 
\begin{figure*}                                                  
\begin{center}
\includegraphics[width=50mm,height=174mm,angle=270]{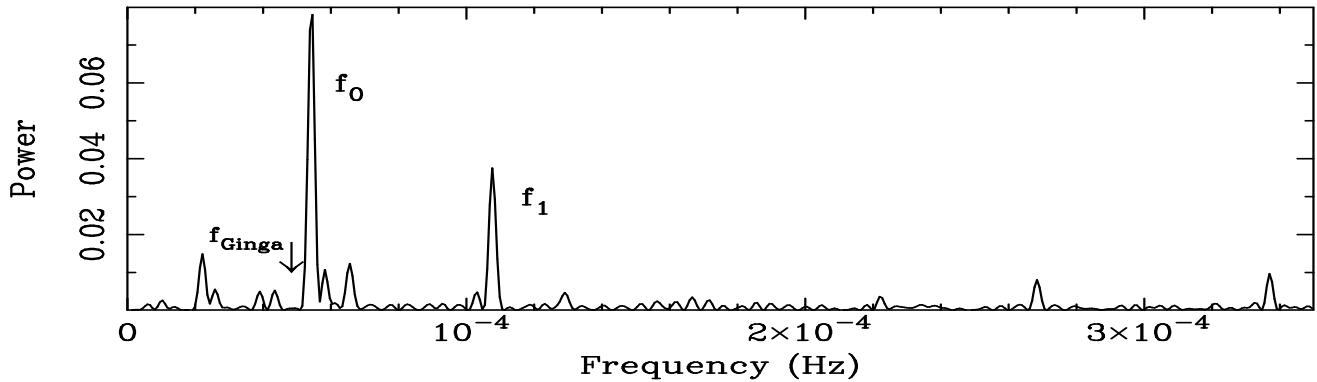}     
\caption{Power spectrum of the 2.1 -- 3.7 keV light curve of the first observation,
after removal of the X-ray bursts, removal of the data gaps and detrending.}
\end{center}
\end{figure*}
\begin{figure*}                                                  
\begin{center}
\includegraphics[width=54mm,height=174mm,angle=270]{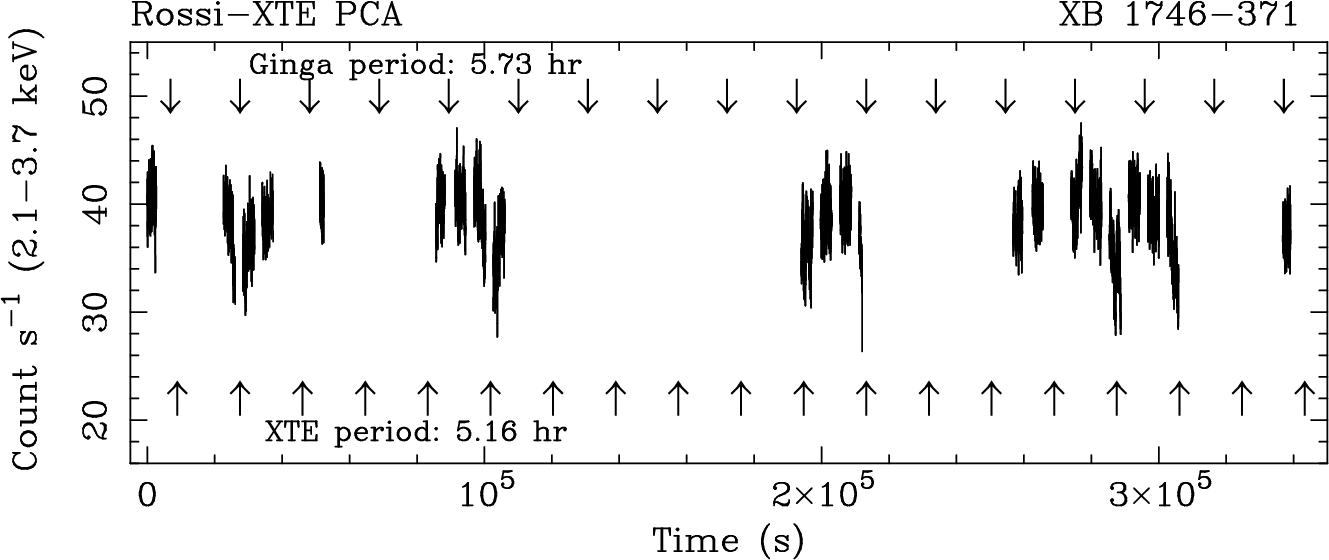}        
\includegraphics[width=54mm,height=174mm,angle=270]{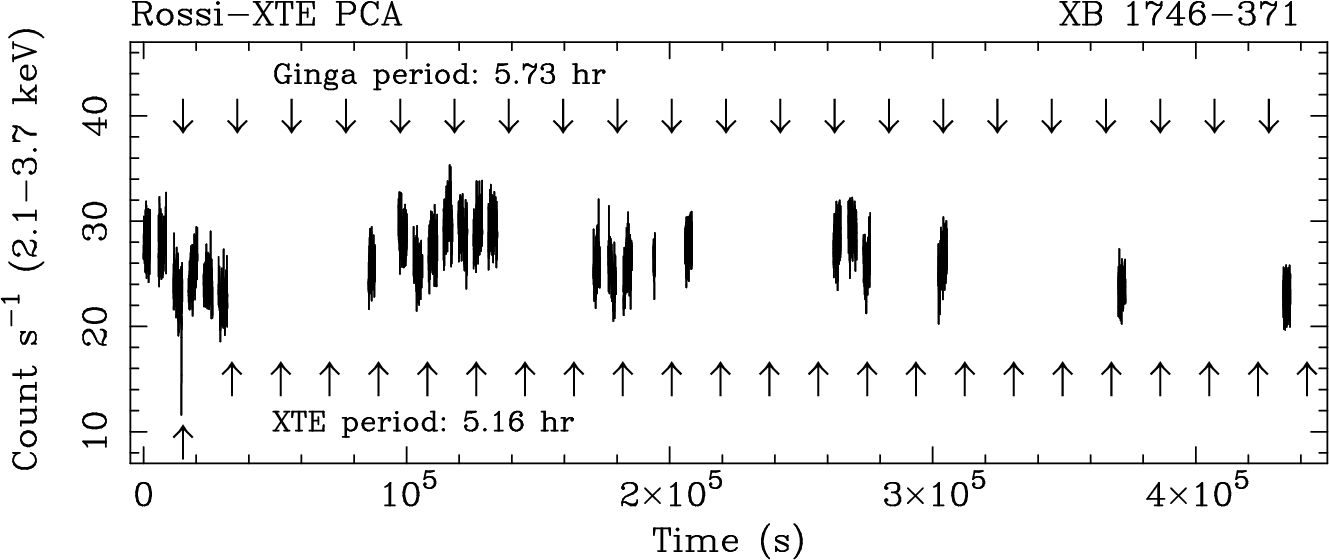}        
\caption{Comparison of the present {\it RXTE} period and the {\it Ginga} period 
with dipping occurrence: upper panel: 2002, Jan observation; lower panel: 2002, May observation.}
\end{center}
\end{figure*}
\begin{figure*}                                                  
\begin{center}
\includegraphics[width=60mm,height=120mm,angle=270]{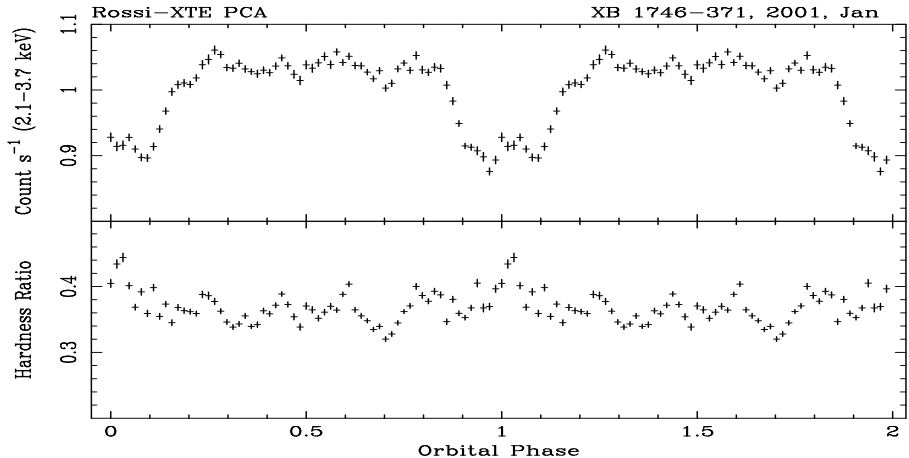}        
\caption{Light curve folded on the orbital period of 5.16 hr obtained in the present work
for the band 2.1 -- 3.7 keV (see text), together with the hardness
ratio (9.8--16.0~keV)/(2.1--3.7~keV). The count rate is normalized by dividing by the mean.}
\end{center}
\end{figure*}
A strong peak is observed
at a frequency ($f_0$ in Fig. 2)
of $5.422\times 10^{-5}$ s$^{-1}$ corresponding to a period of 5.12 hr.
The uncertainty in this value was obtained by the method
of Schwarzenberg-Czerny (1991) using the power spectrum linewidth. This is done by
measuring the power of the noise in the neighbourhood of the main peak and 
obtaining the half-width of the line at a depth of the noise power below the peak
which is the 1$\sigma$ uncertainty. In this case, this gave a half-width in frequency
of $\sim\;1\times 10^{-7}$ Hz . However, this is smaller than the frequency binning in Fig. 2
(optimised in the power spectrum program)
which has a half-width of ${\rm 3.7\times 10^{-7}}$ Hz and so we use this to give
the uncertainty in the period of 5.12$\pm$0.035 hr. 
There is also a peak ($f_1$)
at a frequency twice $f_0$, suggesting that interdipping takes place between the main dips,
so that two dips occur in each orbital cycle. The remaining peaks are all small, 
and most of them are consistent with beating between $f_0$ or $f_1$ and
the frequencies of Earth occultation and SAA passage which are related
to the 90-minute orbital period of the satellite.
The period obtained from {\it Ginga} is indicated by an arrow in
Fig. 2, showing that there is no power at this frequency.

Our period determination was then refined by cycle counting techniques 
using two prior observations of XB\th 1746-371. Light curves were produced for
these observations in the same way as for the data presented above.
The first of these, made in 1996 October, was a long observation
lasting 600~ks during which clear dipping was observed near the start
of the observation at $t$ = 4000 s and near the end at $t$ = 579840
s. Using the 5.12-hr period
we can deduce that 31 cycles elapsed between these dipping episodes,
and based on this cycle count, we obtain a period of 5.16 hr.
It can be seen that the uncertainty of $\pm$0.035 hr accumulates over 31 cycles to 
1.09 hr which is 21 per cent of the cycle. Thus, it is unlikely that
an error has been made in the cycle count.

A second observation made in 1998 consisted of 4 sub-observations
spaced over 6 months. In each of the second and third sub-observations
there was one very well-defined dip with a deep sharp minimum 
(other observed dips were not well-defined). 
The time separation of these dips was 8317920~s, which for 
a period of 5.16 hr gives a cycle count of 447.8. 
The uncertainty in locating dipping due to the time binning of 16 s
gives a negligible error in the period. However, from the three observations
used to refine the period, we estimate a period error of $\sim$0.01 hr.
Using the implied span of periods between 5.15 and 5.17 hr
gives counts of 448.6 and 446.9, respectively, i.e. the count is 448$\pm$1.
Conversely, counts of 449 and 447 correspond to periods of 5.15 and 5.17 hr.
Thus we cannot be certain that there is not an error of one in the count
and so obtain from the above timespan and count a period of 5.15(7)$\pm$0.01 hr.

Finally, we repeated this procedure using the two observations performed in 2002.
The time separation between the lowest intensity point in the last dip in the 
January observation and the lowest point in the first dip in the May observation 
was 8923424 s. Using a period of 5.16 hr, this gives a cycle count of 480.37 for 
this separation. For a cycle count of 480, we obtain a period of 5.16(4) hr. 
Thus, we derive a consistent value of the period using 3 observations
of 5.16$\pm$ 0.01 hr.

This value of the period was then tested against several light curves as follows.
In Fig. 3, we show the present observations and compare the actual
dipping with the predicted
dip centers based on a period of 5.16 hr (lower arrows). We also show
the dip centers expected from the {\it Ginga} period of 5.73 hr (upper arrows). For both periods, the first
arrow is aligned with the first dip in each observation.
It is clear that the
5.16-hr period gives very good alignment with the dips observed. In the second
observation (2002 May), there is a possible dip at $3.0\times 10^5$ s which
is not well-aligned with an arrow. However, the previous dip is at $2.7\times 10^5$ s
which is 7.77 hr from the possible dip. Thus it is likely that this is not a real dip
but part of a trend of general intensity decrease that can be seen in the latter
part of the observation. The 5.16-hr period was also tested against
the light curve of the {\it Exosat}
observation of 1985 September 9,
and gave good alignment.
Finally, the two additional {\it RXTE} observations of 1996 and 1998 used above to refine
the period were tested, and good alignment of dipping with the period obtained.

In Figure 4 we present the data from the 2002 January RXTE observation
folded on the best-fit period, together with
the corresponding folded hardness ratio, where the hardness is defined
as the
ratio of the count rates in the 9.8 -- 16.0 keV and 2.1 -- 3.7 keV bands. The dipping is seen
clearly in the folded light curve, with some indication of
interdipping between the main dips. The folded hardness ratio confirms
that dipping is largely energy independent, as was previously known
from the {\it Exosat} and {\it BeppoSAX} observations (Parmar et
al. 1989, 1999). Because flaring may persist during dipping, we cannot
say whether the small change in hardness in the dip is due to dipping or to
flaring.

\section{Discussion}

The long observations that we made with {\it RXTE} have allowed us to
derive an improved orbital period of 5.16$\pm$0.01~hr for XB\th 1746-371.
Comparison with previous orbital period determinations
shows that the {\it Exosat} value of $5.0\pm 0.5$ hr (Parmar et
al. 1989) is consistent with our present result, as is the {\it
BeppoSAX} period of $5.8^{+0.3}_{-0.9}$ hr (Parmar et al. 1999), but
the {\it Ginga} value of $5.73\pm 0.15$ hr (Sansom et al. 1993) is
not. This is also apparent from Fig. 2 where the {\it
Ginga} value corresponding to a frequency of $4.84\times 10^{-5}$
s$^{-1}$ lies well to the side of the strongest peak. Inspection of
Figs. 3 and 4 also
shows that the {\it Ginga} period is inconsistent with the present data.  In the
2002 January observation, the {\it Ginga} period fails to align with the
dipping at $\sim 10^5$ s, or at $\sim 3\times 10^5$ s, although the
arrows are coincidentally aligned with dipping in between these times
(at $\sim2\times 10^5$ s).  We note that dipping in the {\it Ginga}
data was much less
pronounced than in the present observations, making the period more
difficult to determine.


XB\th 1746-371 is unusual in displaying energy-independent dipping, as does one other
dipping source, X\th 1755-338 (White et al. 1984; Mason et al. 1985;
Church \& Ba\l uci\'nska-Church 1993).
In the case of the {\it Exosat} observation of XB\th 1746-371, 
Parmar et al. (1989) examined possible causes of energy independence, including an ionized
absorber. The {\it BeppoSAX} observation again revealed energy independence in dipping,
and Parmar et al. (1999) ruled out possible causes such as an ionized absorber,
very low metallicity (reduced by 130 times from Solar), and also
the explanation for X\th 1755-338 of Church \& Ba\l uci\'nska-Church (1993)
in which two spectral components combine to give approximate energy independence.
We will investigate the spectral evolution during the dips in the present {\it RXTE} 
data in a further paper.

\vskip 6mm\noindent
{\bf ACKNOWLEDGEMENTS}

\vskip 2 mm\noindent
This research has made use of data obtained from the High
Energy Astrophysics Science Archive Research Center (HEASARC),
provided by NASA's Goddard Space Flight Center.
The work was supported in part by the Polish KBN grants
PBZ-KBN-054/P03/2001 and KBN-2-P03D-015-25.

\end{document}